\documentclass[12pt]{article}
\usepackage{graphicx}
\begin{document}

\begin{center}

{\large \bf Kant and Hegel in Physics}\\[2ex]
Y. S. Kim  \\
Center for Fundamental Physics, University of Maryland,\\
College Park, Maryland 20742, U.S.A.  \\
e–mail: yskim@umd.edu

\end{center}

\vspace{20mm}
\abstract{
Kant and Hegel are among the philosophers who are guiding
the way in which we reason these days.  It is thus of interest to see
how physical theories have been developed along the line of Kant and
Hegel.  Einstein became interested in how things appear to moving
observers.  Quantum mechanics is also an observer-dependent science.
The question then is whether quantum mechanics and relativity can
be synthesized into one science.  The present form of quantum field
theory is a case in point.  This theory however is based on the
algorithm of the scattering matrix where all participating particles
are free in the remote past and in the remote future.  We thus need,
in addition, a Lorentz-covariant theory of bound state which will
address the question of how the hydrogen atom would look to moving
observers.  The question is then whether this Lorentz-covariant
theory of bound states can be synthesized with the field theory into
a Lorentz-covariant quantum mechanics. This article reviews the
progress made along this line. This integrated Kant-Hegel process is
illustrated in terms of the way in which Americans practice their
democracy.
}

\vspace{4cm}

\newpage

\section{Introduction}\label{intro}
\begin{figure}[thb]
\centerline{\includegraphics[width=12cm]{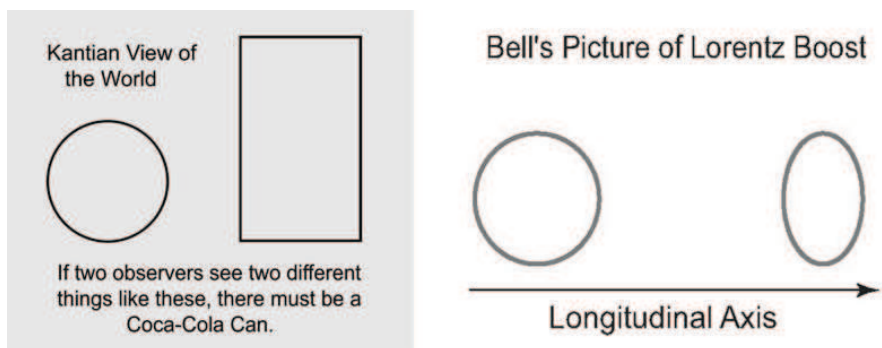}}
\caption{A Coca-Cola can appears differently to two observers from two
different angles. Likewise, the electron orbit in the hydrogen atom should
appear differently to two observers moving with two different speeds.
The elliptic deformation of the circular orbit is from J.S. Bell's
book~\cite{bell04}.  This is deformation is only a speculation based on
Einstein's length contraction and does not have any scientific
merits. The purpose of the present paper is to clarify this
issue.}\label{coke}
\end{figure}

Let us look at a Coca-Cola can.  It is a circle if we see it from its top,
and its side view is a rectangle, as is illustrated in Fig.~\ref{coke}.
How would it then appear to a moving observer.
This is of course an Einsteinian problem.  Einstein studied the philosophy
of Kant during his high school years.  It is thus natural for him to ask
this question.

Niels Bohr was interested in the electron orbit of the hydrogen atom.
It is well known how his efforts led to the present form of quantum
mechanics.  The wave function of the hydrogen atom consists of the
rotation-invariant radial function and the angular function consisting
of spherical harmonics and spinors. The angular function tells how
the orbit looks to observers looking at different angles.  This is
called the rotational symmetry in physics.
The question is how this atom would appear to observers in motion.

\vspace{6mm}

This is a Kantian question.  The observer is in a different environment.
Einstein formulated this problem with a mathematics called the Lorentz
group.  His Lorentzian system consists of three rotations around three
different directions in the $(x,~y,~z)$ space, three Lorentz boosts in
those directions, and three  translations along these different
directions, plus one translation along the time.  There are thus ten
operations in the Lorentzian system.  three translations along the
three different directions.  The mathematics governing these ten
operations is called the inhomogeneous Lorentz group {Wigner 1939].
The main purpose of this paper is to examine how the hydrogen appears
to moving observers in terms of the way Kant and Hegel suggested.
We can then ask whether quantum mechanics and special relativity
can be synthesized.

In Sec.~\ref{kanthegel}, we examine how the idea of Kant and that of
Hegel can be integrated into a single Kant-Hegel procedure in physics.
In Sec.~\ref{synqr}, we review the attempts made in the past to
synthesize quantum mechanics and special relativity.  It is noted that
the present form of quantum field theory can only deal with scattering
problems.  It is noted also that Paul A. M. Dirac made his life-long
efforts to construct bound-state wave functions that can be
Lorentz-boosted.  By integrating those efforts. It is shown possible
to construct harmonic-oscillator wave functions that can be Lorentz-boosted.

In Sec.~\ref{quapar}, we examine whether this Lorentz-transformable
wave function can explain what we see in the real world.  Let us pick
a proton which is a bound state of more fundamental particles called
quarks~\cite{gell64}.  When it moves with a speed close to that of
light, it appears as a collection of free particles called
partons~\cite{feyn69}.
Why does the same proton appear differently?  This is precisely
Einstein's Kantian question.  After settling the issue of bound states
in the Lorentzian system. We are led to the question of whether
quantum mechanics and Einstein's special relativity can be derived
from the same set of mathematical formulas.  It is noted that Dirac
in 1963 started with two harmonic oscillators satisfying the Heisenberg
uncertainty brackets~\cite{dir63}.  He then noted that the symmetry
from these two oscillators is like that of the Lorentz transformations
applicable the five dimensional space with three space coordinate
and two time-like coordinates.

In Sec.~\ref{intqr}, it is shown that the second time variable in this
five dimensional space can be transformed to the translations along the
three space-like directions and one time like direction, just like
Einstein's Lorentzian system.  Indeed, quantum mechanics and special
relativity can be derived from the same set of equations, namely
the Heisenberg brackets.

Kant and Hegel developed their theories based on human societies
and histories, not on physical theories.  It is thus easier to
illustrate the integrated Kant-Hegel mechanism in terms of history.
The history is the United States is short and transparent.  In the
Appendix, we examine the role of Kant and Hegel while Americans
practice their democracy.

\section{Integration of Kant and Hegel}\label{kanthegel}

As is indicated in Fig.~\ref{kanth}, Immanuel Kant and Georg Wilhelm
Friedrich  Hegel are among the most respected philosophers. Yet,
their books are very difficult to read. The best way to understand
their ways of reasoning is to construct illustrations.

According to Kant, many things should become one, the ding-an-sich.
They just look differently depending on the observer's environment
and state of mind.  According to Hegel, we can create a new wonderful
world by synthesizing two different traditions.  His philosophy was
based on the history.  He realized that Christianity is a synthesis
of Jewish ethics and Greek philosophy.  How can we integrate Kant and
Hegel?  Kant wanted to derive one from many.  Hegel wanted to derive
one from two.

Thus, we need a mechanism which will lead many to two, between Kant
and Hegel.  This way of thinking was developed by ancient Chinese.
After the ice age, many people with different backgrounds came to
the banks to China's northern river.  They drew pictures to
communicate, and this led to Chinese characters.  In order to express
their feelings, they sang songs.  This is the reason why there are
tones in spoken Chinese. How about different ideas?  They realized
they cannot be united to one.  They thus divided them into two opposing
groups, namely Yang (plus) and Ying (minus).  This way is known as
Taoism.

\begin{figure}
\centerline{\includegraphics[width=12cm]{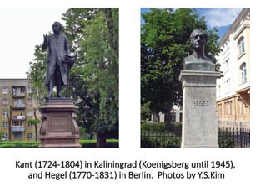}}
\caption{Kant and Hegel.  Einstein had a very strong Kantian background
from his high school years.  Yet, he became a great Hegelian synthesizer
as a physicist.  His photo-electric effect synthesized the particle and
wave natures of matter.  He synthesized the energy-momentum relations for
massive and massless particles.  This work is commonly known as Einstein's
$E = mc^2$.}\label{kanth}
\end{figure}

\begin{figure}
\centerline{\includegraphics[width=12cm]{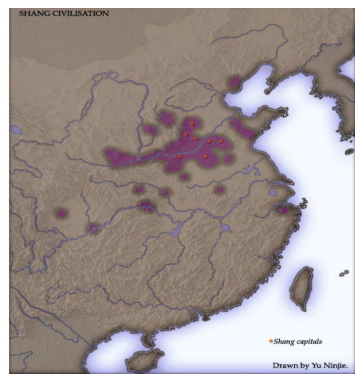}}
\caption{Geography of Kant's East Prussia and Geography of ancient China.
Different groups with their own ways of thinking come to the same area,
Their views toward the same thing could be different.}\label{tao}
\end{figure}

Kant was born in East Prussia (now Kaliningrad, Russia), and spent
80 years of his entire life there.  Thus, his way of thinking was
framed by what he saw every day.  His area was a maritime commercial
hub of the Baltic Sea, just like Venice in the Mediterranean world.
Many people came to Kant's place with many different points of view
for the same thing. Thus, Kantianism and Taoism were developed in
the same way, as illustrated in
Fig.~\ref{tao}.  Kant wanted one, but Chinese had to settle with two.
Thus, Taoism can stand between Kant and Hegel.  We can thus integrate
Kant and Hegel by placing Taoism between them.
We can illustrate this integrated Kant-Hegel system in terms
of the history of the United States.  Europeans with different
backgrounds came to the new land and settled down in many different
areas.  They then set up their own government.  In order to develop
their laws and national policies, they developed two different
political parties. These two parties produce the laws and policies
applicable to all citizens. This American system is admired by many
people of the world.  This is an integrated Kant-Hegel system, as
is illustrated in Fig.~\ref{kantao}.

\begin{figure}
\centerline{\includegraphics[width=11cm]{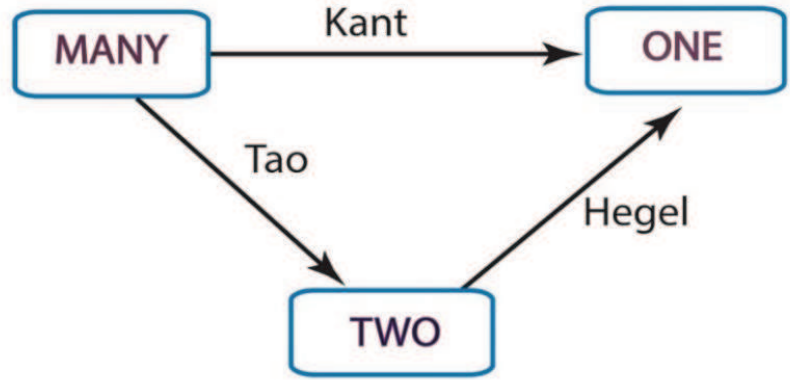}}
\caption{Integration of Kantianism and Hegelianism.
Kant wanted to reduce many-to-one while Hegel wanted to
combine two-into-one.  Thus, we need many-to-two mechanism
in between.  This is what Taoism is about.}\label{kantao}
\end{figure}

In the past, the physical laws were developed according to this
integrated Kant-Hegel process.  Physicists like to unite many different
events into one formula or one set of formulas.
There are many heavenly bodies.  They can be divided into two groups,
namely comets (with open orbits) and planets with (localized orbits).
Isaac Newton synthesised these two groups into one with his second-order
differential equation.  This is a Hegelian process.
James Clerk Maxwell synthesised the equations governing electricity
and those for magnetism into a set of four equations.  This created
the present-day wireless civilization.
As is well known, the present form of quantum mechanics is a synthesis
of particle nature and wave nature of matter.  Einstein's special
relativity synthesizes massive and massless particles.  All these are
the integrated processes of Kantianism and Hegelianism.
The remaining problem is whether quantum mechanics and Einstein's
relativity can be synthesised.  In this paper, we restrict ourselves
to his special relativity, even though his general relativity receives
more public attention these days.  Einstein's nickname is still
$E = mc^2$, which was a product of his special theory of relativity.

\section{Synthesis of of Quantum Mechanics amd Special Relativity}\label{synqr}
The present form of quantum mechanics was developed for the Galilean
system. While the Galilean system operates with three translations and
three-rotations on the space of $(x,~y,~z)$, the time variable does not
interfere with the coordinate transformations. However, was stated in
Sec.~\ref{intro}, the space and time of Einstein's special
relativity is based on the Lorentzian system.  This system operates in
the four-dimensional space of $(x,~y,~z)$.   In this Lorentzian system,
there are rotations in the three-dimensional space of $(x,~y,~z)$.
In addition, when the observer moves with a constant speed, the time
variable comes in.  We call the observer's velocity change ``Lorentz
boost.''  The boost can be made in three different directions.  We call
this symmetry system Lorentz covariance.  In additional, there are four
translations along the four-dimensional space.  Let us call the system
of these three rotations, three boosts, and four translation the
Lorentzian system.

\begin{table}[thb]
\caption{Three different systems.  In the traditional Galilean system,
there are three rotational and three translational degrees of
freedom.  In the Lorentzian system, four boost operations are
possible and one additional translation, namely along the time
direction. In 1963, Dirac constructed a space-time symmetry from
two harmonic oscillators~\cite{dir63}. This will be discussed
later in this paper.}\label{tab11}
\vspace{0.5mm}
\begin{center}
\begin{tabular}{lcccccc}
\hline
\hline\\[-0.4ex]
\hspace{1mm}
Systems &{}& \hspace{3mm} Galilean & \hspace{3mm} &
     Lorentzian &\hspace{3mm}& Two Oscillators  \\[1.0ex]
\hline\\ [-0.7ex]
\hspace{1mm}
Rotations &{}&  $ J_{x}, J_{y}, J_{z}. $
   &\hspace{10mm}&
   $ J_{x}, J_{y}, J_{z}. $ &{}&  $ J_{x}, J_{y}, J_{z}. $
\\[2ex]
\hline\\ [-0.7ex]
\hspace{1mm}
Boosts &{}& None
   &\hspace{10mm}&
   $ K_{x}, K_{y}, K_{z}. $ &{}&  $ K_{x}, K_{y}, K_{z}. $
\\[2ex]
\hline\\ [-0.7ex]
\hspace{1mm}
Translations &{}&
$ P_{x}, P_{y}, P_{z}. $ &{}&  $ P_{x}, P_{y}, P_{z}, P_{t}. $
 &{}&
$ \matrix{ Q_{x}, Q_{y}, Q_{z}, S_{0} \cr \mbox{contracted to} \cr
   P_{x}, P_{y}, P_{z}, P_{t}. } $
\\[5ex]
\hline
\hline\\[-0.4ex]
\end{tabular}
\end{center}
\end{table}

The difference between the Galilean system and the Lorentzian
system is spelled out in Table~\ref{tab11}.  Quantum mechanics was
originally developed in the Galilean system, but it was a great
challenge during the 20th Century to construct quantum mechanics
in the Lorentzian space and time.

\begin{figure}
\centerline{\includegraphics[width=12cm]{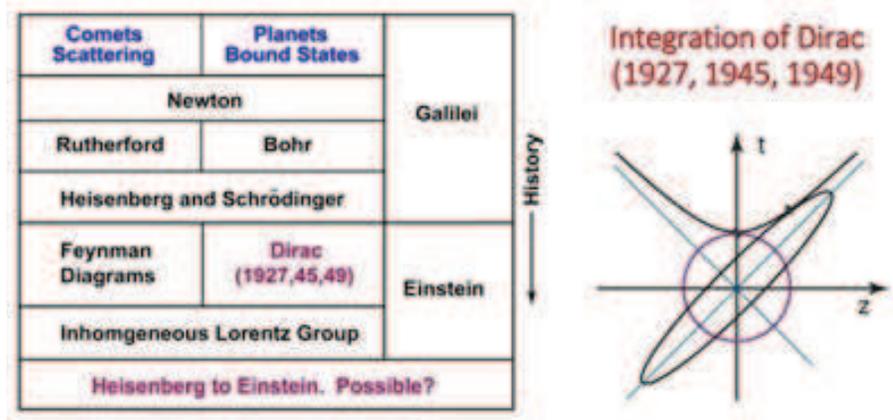}}
\caption{Historically, our unified understanding of open and closed
orbits has been very brief.  Quantum field theory and covariant
oscillator lead to quite different mathematical formulas. However,
they both are representations of the inhomogeneous Lorentz
group~\cite{kno79}.}\label{comet}
\end{figure}

Quantum field theory is a case in point.  The mathematical algorithm
of this theory is based on the scattering matrix where all participating
particles are free in the remote past and free in the remote future.  For
 making computations of the scattering matrix, Feynman diagrams provide
 mathematical transparencies with excellent physical interpretations.
How about bound-state particles?  The particles are not free in the
remote past and remote future.

As indicated in Fig.~\ref{comet}, our understanding of bound states
and scattering states did not go together all the time.  As Feynman
suggested~\cite{fkr71}, while Feynman diagrams are useful for running
waves, we can use harmonic oscillators to understand quantum bound
states in the Lorentzian system.  In their paper of 1971, however,
Feynman {\it et al.} did not do a very good job in constructing the
harmonic oscillators in the Lorentzian world.  Their oscillator wave
functions increase as the time separation variable become large.  Thus,
their wave functions are meaningless in quantum mechanics Indeed,
before Feynman {\it et al.}, Dirac attempted to construct a
representation of the Lorentz group using harmonic oscillator wave
functions~\cite{dir45}.  Before 1971, a number of authors wrote down
Lorentz-covariant oscillator wave
 functions~\cite{yuka53,markov56,fuji70}.

Yet, Feynman {\it et al.} ignored them all. Since 1973, mostly with
Marilyn Noz, the present author started publishing papers on this
subject~\cite{kn73} and continued writing papers and books
along the same line~\cite{kno79,kn77,knp86,kim89,bkn19qr,kn20sym}.
With those papers, it is now possible to integrate Dirac's lifelong
efforts to construct Lorentz-covariant oscillator wave functions.
Dirac published three papers toward the Lorentz-covariant oscillators.
In 1927, Dirac said that the c-number time-energy uncertainty should
be included in Einstein's Lorentzian world~\cite{dir27}.  In 1945,
he suggested harmonic oscillators for a representation of the Lorentz
group~\cite{dir45}. In 1949, he introduced the light-cone coordinate
system for Lorentz boosts, saying that the Lorentz boost is a squeeze
transformation~{dir49}.  Dirac's papers are like poems, but they
contain no diagrams.  Thus, we can use diagrams to accomplish what
Dirac did not do, that is to integrate his own papers.  As is
illustrated in Fig.~\ref{truck}, his three papers can be integrated
into an ellipse as a squeezed circle tangent to Einstein's hyperbola.

\begin{figure}
\centerline{\includegraphics[width=12cm]{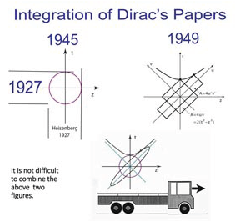}}
\caption{ Translation of Dirac's papers into pictorial language.
The synthesis of his 1927 and 1945 result in a circle.  His squeeze
 transformation of 1949 leads to a squeezed circle or ellipse shown
 in this figure.}\label{truck}
\end{figure}

Let us go back to Fig.~\ref{comet}, it is important to note
that both the Feynman diagrams and the oscillator formalism given
in this section can be constructed from the same set of commutation
relations, which is known as the Lie algebra of the inhomogeneous
Lorentz group~\cite{kno79}.  It is also important note
that the Feynman diagrams and the Lorentz-covariant oscillator
wave functions are constructed from the same set of physical
principles governing quantum mechanics and special
relativity~\cite{hkn81fp}.

\section{Quark-parton Puzzle}\label{quapar}
On hundred years ago, Bohr and Einstein met occasionally to
discuss physics.  Bohr was worrying about the electron orbit of
the hydrogen atom, while Einstein's main interest was how things
appear to moving observers. Thus, they could have talked about
how the hydrogen atom looks to a moving observer.  However,
there are no records indicating that they ever talked about
this issue.  If they did not, they are excused.  There were
and still are no hydrogen atoms moving with relativistic speeds.
Since the total charge of the hydrogen atom is zero, it cannot
be accelerated even these days.

\begin{figure}
\centerline{\includegraphics[width=12cm]{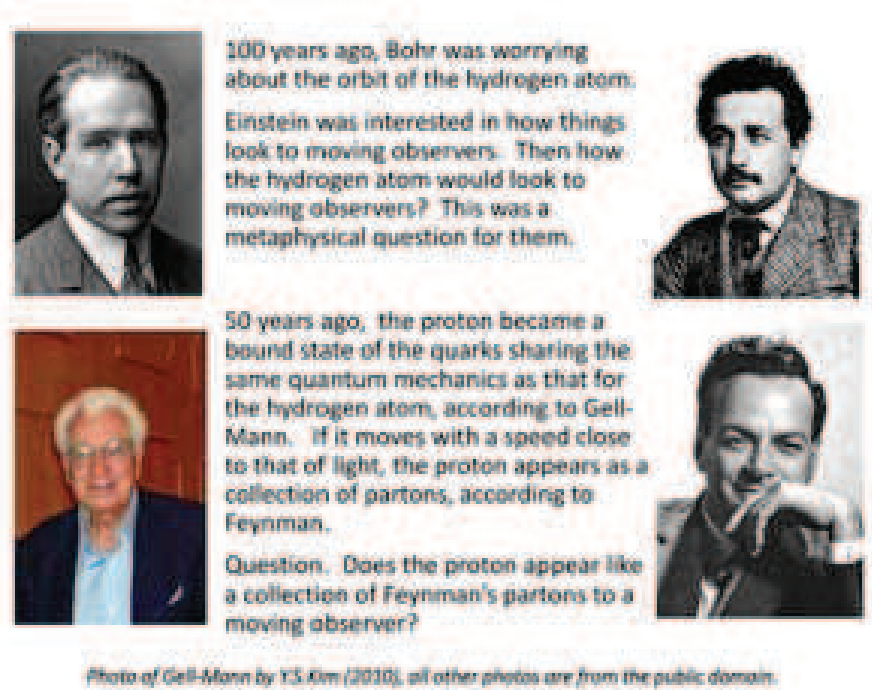}}
\caption{The Bohr-Einstein on the hydrogen, and Gell-Mann and
Feynman on the proton. Both the hydrogen and proton are bound
states sharing the same quantum mechanics.  The proton can be
accelerated, while this is not the case for hydrogen
atom.}\label{einbo}
\end{figure}

On the other hand, modern particle accelerators routinely produce many
protons moving with the speed very close to that of light.  These
protons are not hydrogen atoms.   However, they are also bound states
within the same framework of quantum mechanics.
 As indicated in Fig.~\ref{einbo}, it is possible to study moving
 hydrogen atoms by looking at moving protons.
Indeed, according to the quark model~\cite{gell64}, the proton
is a quantum bound state of three quarks.  Then the question is how
the proton appears when it moves fast.  In 1969, Feynman observed
that the proton looks quite differently when it moves with
ultra-fast speed~\cite{feyn69}.  It appears like a collection of
light-like particles.  Feynman called them partons.  These partons
have the following peculiar properties.

\begin{itemize}
\item[a.] Feynman's parton picture is valid only for protons moving
with velocity   close to that of light.
\item[b.] The interaction time between the quarks becomes dilated,
and partons behave like free independent particles.
\item[c.] The momentum distribution of partons becomes widespread
as the proton speed increases.
\item[d.] The number of partons seems to be infinite or much larger
than that of the constituent quarks.
\end{itemize}

\begin{figure}
\centerline{\includegraphics[width=12cm]{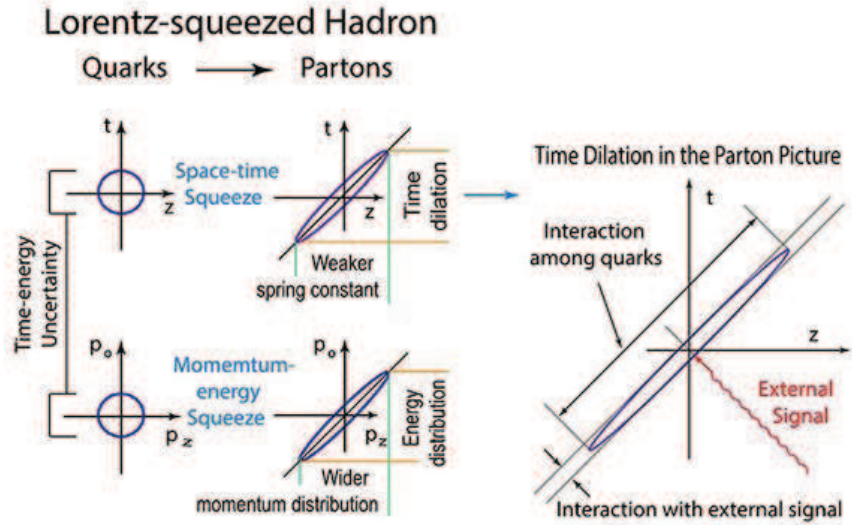}}
\caption{Space-time and momentum energy wave functions.  As the
hadronic speed reaches that of light, both wave functions become
concentrated along one of the light cones.}\label{parton}
\end{figure}

The question is whether it is possible to explain Feynman's parton
picture within the framework of quantum mechanics and special
relativity.  In order to answer this question, we need a bound-state
wave function which can be Lorentz-boosted. In Sec.~\label{synqr},
we constructed the Gaussian function that can be Lorentz-boosted.
The Gaussian function is different from the wave
function for the hydrogen atom, but both wave functions share the
same quantum mechanics.  In 1964, Gell-Mann proposed the quark model
for hadrons.  The hadrons are bound states of the quarks, and their
mass spectra are like those of the harmonic
oscillators~\cite{fkr71}.  The proton is a hadron and is
a bound state of three quarks.  In the oscillator regime, the wave
function for this three-body system is a product of two wave
functions~\cite{fkr71}.  It is thus sufficient to study the
Lorentz-boost property of the two-quark system which was discussed
in Sec.~\ref{quapar}.  In the oscillator system, the momentum
wave function of the Gaussian wave function is also Gaussian,
and it becomes Lorentz-squeezed exactly in the same way as in
the case of the space-time wave function.  Thus, we can extend
Fig.~\ref{truck} to Fig.~\ref{parton}~\cite{kn77,kim89}.

Indeed, according to this figure, the quarks become light-like
particles with a wide-spread momentum distribution, interacting with
an external like free particles.  Furthermore, since they are like
light-like, the particle number is not constant as in the case of
black-body radiation.  Thus, this figure provides the answers to
all of the puzzles raised on the parton picture listed earlier in
this section.  As was stated before, the proton wave function is a
product of two oscillator wave functions.  Using this wave function,
Paul Hussar computed the parton distribution function for the proton,
and it is in reasonable agreement with the observed parton
distribution~\cite{hussar86}.  Let us go to Table~\ref{tab22}.  The
Lorentz-covariant oscillator, which can be viewed as an integration
of Dirac's papers, provides a one formula for the proton at rest
(quark model) and the ultra-fast proton (parton picture).  The second
row of this table is about the role of Wigner's little group for internal
space-time symmetries~\cite{wig39}. This row tells that Wigner's little
groups explain why the internal space-time symmetries appear differently
to moving observers.  Again, this is a Kantian problem and was discussed
detail in the literature~\cite{kn77,hks83pl,kim89,kiwi90jmp,kn18iop}.

\begin{table}
\caption{Further contents of Einstein's $E = mc^2$.  The first row
of this table in well known.  The second table is for internal
space-time symmetries, the photon spin along the direction of momentum
remains invariant, but the perpendicular components become one gauge
degree of freedom~{\it et~al.}~1983, Ki~and~Wigner~1990].
The covariant oscillator explains the peculiarities of Feynman's
parton picture}\label{tab22}
\vspace{2mm}
\begin{center}
\begin{tabular}{lcccccc}
\hline
\hline\\[-0.4ex]
\hspace{1mm}
{} &{}& \hspace{3mm} $\matrix{\mbox{Slow} \cr \mbox{Masssive}}$  & \hspace{3mm} &
between  &\hspace{3mm}& $\matrix{\mbox{Ultra fast} \cr \mbox{Massless}} $\\[1.0ex]
\hline\\ [-0.7ex]
\hspace{1mm}
$\matrix{\mbox{Einstein's} \cr E = mc^2}$ &{}&  $ E = p^2/2m $
   &\hspace{10mm}&
   $ E = \sqrt{(cp)^2 + m^2c^4} $ &{}&  $ E = cp $
\\[2ex]
\hline\\ [-0.7ex]
\hspace{1mm}
$\matrix{\mbox{Wigner's} \cr \mbox{Little Groups}} $ &{}& $\matrix{ S_3, \cr S_1, S_2}$
 &\hspace{10mm}& $\matrix{\mbox{Internal} \cr \mbox{Symmetry}  } $
    &{}& $\matrix{\mbox{Helicity} \cr \mbox{Gauge Trans.}}$
\\[2ex]
\hline\\ [-0.7ex]
\hspace{1mm} Integration of &{}& Gell-Mann's &{}& Covariant &{}& Feyman's \\[0.3ex]
\hspace{1mm} Dirac 1927,45,49.  &{}&  Qaurks Model  &{}&  Oscillators
 &{}& Parton Picture
\\[3ex]
\hline
\hline\\[-0.4ex]
\end{tabular}
\end{center}
\end{table}

\section{Integration of Quantum Mechanics and Special Relativity}\label{intqr}

In Secs. 4 and 5, we observed that it is possible to construct
harmonic oscillator wave functions that can be Lorentz-boosted.
Furthermore, it can settle the one of the Kantian problems we
observe in high-energy laboratories producing ultra-fast  protons.
In addition, we noted that the covariant oscillators and quantum
field theory can be constructed within the same Lorentzian system.
They also share the same set of physical of physical principles.

In that case, we are led to the question of whether these two
scientific disciplines can be derived from the same set of equations.
For this purpose, Dirac considered two harmonic oscillators. For the
single-oscillator system, we can use the step-up and step-down operators
to write down Heisenberg's uncertainty brackets.

\begin{table}
\caption{Dirac's ten quadratic forms which satisfy a closed set of
commutation relations identical to that of the Lorentz group applicable
to three space-like coordinates and two time-like coordinates.  The J
operators are for the rotations in the three-dimensional space. The three
K operators generate Lorentz boosts along three different directions.
This table contains only six of the ten generators.}\label{tab33}
\vspace{3mm}
\begin{center}
\begin{tabular}{llllll}
\hline
\hline\\[-0.4ex]
\hspace{1mm}& Dirac Oscillators  &{}& \hspace{3mm} Differential   \\[0.8ex]
\hline\\
\hspace{1mm}&
$J_{1} = {1\over 2}\left(a^{\dag}_{1}a_{2} + a^{\dag}_{2}a_{1}\right)$ &{}&
$ -i\left(y\frac{\partial}{\partial z} - z\frac{\partial}{\partial y}\right) $
\\[2ex]
\hline\\
\hspace{1mm}&
 $J_{2} = {1\over 2i}\left(a^{\dag}_{1}a_{2} - a^{\dag}_{2}a_{1}\right) $ &{}&
$  -i\left(z\frac{\partial}{\partial x} - x\frac{\partial}{\partial z}\right)$
\\[2ex]
\hline\\ [-0.7ex]
\hspace{1mm}  &
$J_{3} = {1\over 2}\left(a^{\dag}_{1}a_{1} - a^{\dag}_{2}a_{2} \right)$ &{}&
$ -i\left(x\frac{\partial}{\partial y} - y\frac{\partial}{\partial x}\right)$
\\[2ex]
\hline\\ [-0,7ex]
\hspace{1mm}  &
$K_{1} = -{1\over 4}\left(a^{\dag}_{1}a^{\dag}_{1} + a_{1}a_{1} -
  a^{\dag}_{2}a^{\dag}_{2} - a_{2}a_{2}\right) $ &{}&
 $ -i\left(x\frac{\partial}{\partial t} + t\frac{\partial}{\partial x}\right)$
\\[2ex]
\hline\\ [-0.7ex]
\hspace{1mm}  &
$K_{2} = +{i\over 4}\left(a^{\dag}_{1}a^{\dag}_{1} - a_{1}a_{1} +
  a^{\dag}_{2}a^{\dag}_{2} - a_{2}a_{2}\right)$ &{}&
$ -i\left(y\frac{\partial}{\partial t} + t\frac{\partial}{\partial y}\right)$
\\[2ex]
\hline\\ [-0.7ex]
\hspace{1mm}  &
$K_{3} = {1\over 2}\left(a^{\dag}_{1}a^{\dag}_{2} + a_{1}a_{2}\right)$ &{}&
$ -i\left(z\frac{\partial}{\partial t} + t\frac{\partial}{\partial z}\right)$
\\[2ex]
\hline
\hline\\[-0.4ex]
\end{tabular}
\end{center}
\end{table}

For the two-oscillator system, there are four such operators. Dirac
constructed ten quadratic forms with those step-up and step-down operators.
He then noted that they satisfy the closed set of commutation relations
which is the same as that for the generators of the Lorentz group
applicable to the five-dimensional space consisting of three space
coordinates $(x,~y,~z)$ and two time coordinates $t$ and $s$~\cite{dir63}.

Table~\ref{tab33} gives three rotation generators and three boost
generators with respect to the time variable t, along with corresponding
Dirac's two-oscillator forms.  The operators of this table do not depend
on the second time variable s. The J operators there generate rotations
in the three-dimensional space $(x,~y,~z)$, and the K operators generate
Lorentz boosts along those three different directions.  Thus, the six
operators given in Table generate the Lorentz group familiar to us.

\begin{table}
\caption{
 Four additional quadratic forms in Dirac's paper of 1963. They correspond
 to the generators operating on the second time variable.}\label{tab44}
\vspace{0.5mm}
\begin{center}
\begin{tabular}{llllll}
\hline
\hline\\[-0.4ex]
\hspace{1mm} Dirac Ocillators &\hspace{10mm} & \hspace{3mm} Differential\\[0.8ex]
\hline\\ [-0.7ex]
\hspace{1mm}
$ Q_{1} = -{i\over 4}\left(a^{\dag}_{1}a^{\dag}_{1} - a_{1}a_{1} -
  a^{\dag}_{2}a^{\dag}_{2} + a_{2}a_{2} \right)$ & {}&
    $ -i\left(x\frac{\partial}{\partial s} +
     s\frac{\partial}{\partial x}\right) $
\\[2ex]
\hline\\ [-0.7ex]
\hspace{1mm}
$ Q_{2} = -{1\over 4}\left(a^{\dag}_{1}a^{\dag}_{1} + a_{1}a_{1} +
   a^{\dag}_{2}a^{\dag}_{2} + a_{2}a_{2} \right)$ &{}&
$  -i\left(y\frac{\partial}{\partial s} + s\frac{\partial}{\partial y}\right)$
\\[2ex]
\hline\\ [-0.7ex]
\hspace{1mm}
$ Q_{3} = \frac{i}{2}\left(a_{1}^{\dag}a_{2}^{\dag} - a_{1}a_{2}\right)$ &{}&
$ -i\left(z\frac{\partial}{\partial s} + s\frac{\partial}{\partial z}\right)$
\\[2ex]
\hline\\ [-0.7ex]
\hspace{1mm} $S_{0} = {1\over 2}\left(a^{\dag}_{1}a_{1} + a_{2}a^{\dag}_{2}\right)$&{}&
$ -i\left(t\frac{\partial}{\partial s} - s\frac{\partial}{\partial t}\right)$
\\[2ex]
\hline
\hline\\[-0.4ex]
\end{tabular}
\end{center}
\end{table}

In addition to the six quadratic forms given in Table~\ref{tab44},
Dirac constructed four additional quadratic forms. They correspond
to the differential operators given in Table~\ref{tab44}.  The
differential operators in Table 4 do not depend on the time variable
$t$, but depend only on the second time variable $s$. We are now
interested in converting the differential forms in this table into
four translation generators, using the group contraction procedure
introduced first by In{\"o}n{\"u} and Wigner~{inonu53}. In their
original paper, In{\"o}n{\"u} and Wigner obtained the Galilean system
from the Lorentzian system. Since then, this contraction procedure was
used for the unification of Wigner's little groups for massive and
massless particles~\cite{hks83pl,kim89,kiwi90jmp}.  This procedure
unifies the internal space-time symmetries of massive and massless
particles, as Einstein's $E = mc^2$ does for the energy-momentum
relation, as indicated in Table~\ref{tab22}. This contraction procedure
has been employed for the present purpose of converting all four
operators in Table~\ref{tab33} into four translation
generators~\cite{bkn19qr,kn20sym}. The result is shown in
Table~\ref{tab55}.

\begin{table}
\caption{Contraction of the s-dependent operators.  According to
In{\"o}n{\"u}-Wigner procedure for group contractions, we can let s = 1.
This procedure transforms
the Q and S operators into translation operators.}\label{tab55}
\vspace{0.5mm}
\begin{center}
\begin{tabular}{lccccc}
\hline
\hline\\[-0.4ex]
\hspace{1mm}{}&{}  & Differential &{}& contracted to\\[0.8ex]
\hline\\ [-0.7ex]
\hspace{1mm}
$ Q_{1}  $ & {}&
      $ -i\left(x\frac{\partial}{\partial s} +
     s\frac{\partial}{\partial x}\right) $
     &{}& $   -i\frac{\partial}{\partial x} $
\\[2ex]
\hline\\ [-0.7ex]
\hspace{1mm}
$ Q_{2} $ &{}&
$  -i\left(y\frac{\partial}{\partial s} + s\frac{\partial}{\partial y}\right)$
   &{}& $   -i\frac{\partial}{\partial y} $
\\[2ex]
\hline\\ [-0.7ex]
\hspace{1mm}
$ Q_{3}$ &{}&
$ -i\left(z\frac{\partial}{\partial s} + s\frac{\partial}{\partial z}\right)$
&{}&$   -i\frac{\partial}{\partial z} $
\\[2ex]
\hline\\ [-0.7ex]
\hspace{1mm}
$S_{0}$ &{}&
$ -i\left(t\frac{\partial}{\partial}  - s\frac{\partial}{\partial t}\right) $
&{}& $ i\frac{\partial}{\partial t} $
\\[2ex]
\hline
\hline\\[-0.4ex]
\end{tabular}
\end{center}
\end{table}

This contraction procedure tells us to fix the s variable and set
$s = 1$ for the generators given in Table~\ref{tab44}~\cite{bkn19qr,kn20sym}.
They then become contracted to the translation generators given in
Table~\ref{tab55}.  Indeed, the six generators of the Lorentz group
given in Table~\ref{tab11} together
with the four translation generators constitute the generators
of the inhomogeneous Lorentz group or Einstein's Lorentzian system
of space and time. This process is compared with the traditional
Galilean and Lorentz systems in Table~\ref{tab11}.

Let us go back to Fig.~\ref{comet}.  This figure asks whether it
is possible to derive Einstein's Lorentzian world from the
principles of quantum mechanics.  The answer to this question is
YES.  While the Dirac's oscillator algebra is derivable from the
Heisenberg brackets, the Heisenberg brackets are also derivable
from the oscillator algebra. Thus, both quantum mechanics and
special relativity are derivable from the same set of equations.
The synthesis of quantum mechanics and special relativity is now
complete.

\section*{Concluding Remarks}
Kant and Hegel are very familiar names to us.  They formulated
their ideas based on what they observed and what they learned.  It
is interesting to note that Einstein started as a Kantianist but
become a Hegelianist while doing physics.  Indeed, physics develops
along the integrated Kant-Hegel line.  The most pressing task of
our time in physics is a Hegelian synthesis of quantum mechanics
and theories of relativity.  For the single-oscillator system, we
can use step-up and step-down operators to write down Heisenberg's
uncertainty brackets.  For the two-oscillator system, there are
four such operators. Dirac constructed ten quadratic forms with
those step-up and step-down operators.  He then noted that they
satisfy the closed set of commutation relations which is the
same as that generators of the Lorentz group applicable to the
five-dimensional space consisting of three space coordinates
$(x,~y,~z)$ and two time coordinates $t$ and $s$~\cite{dir63}.
In this paper, it is pointed out that those ten generators can be
transformed to the ten generators of Einstein's Lorentzian system.
Thus, quantum mechanics and special relativity come from the same
set of equations.

\begin{figure}
\centerline{\includegraphics[width=12cm]{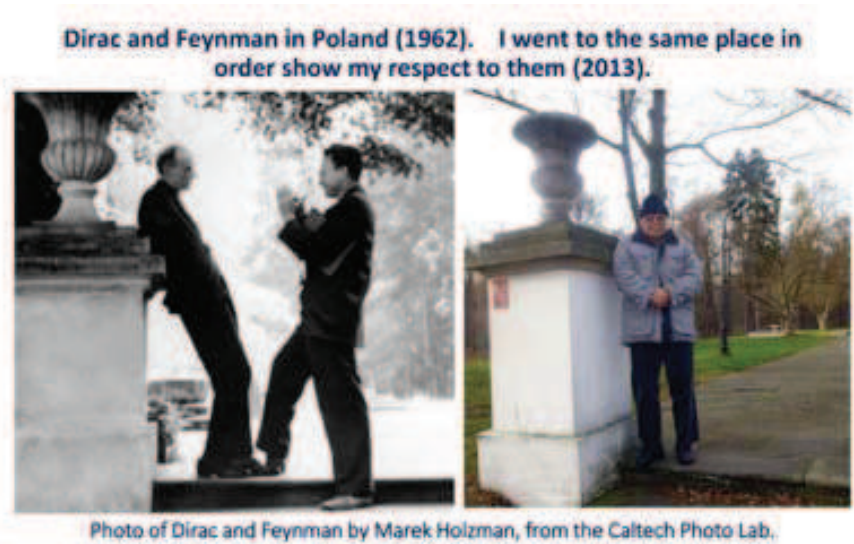}}
\caption{This photo of Dirac and Feynman appeared on the cover
of the Physics Today (August 1963).
This photo was taken during the Relativity Conference organized
by Leopold Infeld in July of 1962 at the Jablonna Palace near
Warsaw, Poland. To show my great respect for both Dirac and
Feynman, I went to the same place and produced my own photo.
I am 168 cm tall and not short, but both Dirac and Feynman had
longer legs.}\label{jablo}
\end{figure}

\section*{Anowledgnents}
I came to the University of Maryland in 1962 as an assistant professor
one year after I received my PhD degree from Princeton in 1961.
At that time, the chairman of the physics department was John S. Tall.
He invited Paul A. M. Dirac to his department for ten days in October
of 1962, and he assigned me to be a personal assistant to Dirac.  I
asked Dirac many questions, but Dirac's answer was very consistent.
American physicists should study more about Lorentz covariance and its
difference from the Lorentz invariance.  Why was he saying this?

Three months earlier (in July 1962), Dirac met Feynman in Poland
during a relativity conference organized by Leopold Infeld. The photo
of their meeting was later published on the cover of the Physics Today,
as shown in Fig.~\ref{comet}.  After reading the 1971 paper by Feynman
{\it et al.}~\cite{fkr71}, it became clear to me that Feynman and his
younger co-authors did not understand the difference between the
covariance and invariance in the Lorentzian world.  This is the reason
why they ignored the Lorentz-covariant oscillator wave functions which
existed in the literature before 1971. When Dirac was telling me about
the weakness of American physicists in 1962, he was talking about
Feynman he met three months earlier in Poland.

Yet, the names of Feynman and Dirac are prominently displayed in
the history table given in Fig.~\ref{comet}. They have been and
still are my great inspirational figures to, and I have been eager
to place them into one box according to the Hegelian process of
synthesis.  In order to show my gratitude toward them, I visited
in 2013 the Jabllona Palace north of Warsaw where they met in 1962,
as shown in Fig.~\ref{jablo}. In had the pleasure of my photo
taken at the spot where they spoke to each other.  Their photo
a lack of communication between them. It has been a great challenge
for me to fix the gap between them.
Finally, I am indebted to John S. Toll who provided my meeting with
Dirac in 1962.  He was always helpful to me whenever I whenever
needed help throughout my academic career.

\section*{Appendix}

Traditionally, philosophers wrote their theories based on the religion,
history, cultural conflicts.  Their theories are quite separate from
physical phenomena.  Thus, it is much easier to illustrate their
philosophies using historical developments.

The history of the United States is short and transparent.  After the
first journey of Christopher Columbus (1492-93), many Europeans moved
to the New Land. In 1776, the Declaration of Independence was ratified.
This document was written before Kant and Hegel became prominent, and
it does not say anything about political parties.

These days, the democratic system of the United States is functioning
with two political parties.  Americans did not construct this system
based on any theories of government written before.  They developed
this two-party system while practicing their democracy.  How did they
construct?  The country consisted of many different ethnic groups with
different cultural backgrounds. They were spread over many different
areas in the North American continent.  How it is possible to construct
a national policy satisfactory to all those citizens?

While practicing democracy, it is necessary to construct one national
policy based on all different opinions.  Thus, the Kantian process of
"many-to-one" is desirable.  However, it is not practical.  Therefore,
a more practical solution was to place those many opinions into two
different groups.  It is then possible to "synthesize" two opinions
into one, according to the Hegelian synthesis of "two-to-one."  This
process is illustrated in Fig.~\ref{kantao}, which illustrates how
physical theories are developed.

\end{document}